\documentclass[journal=jcisd8,manuscript=article]{achemso}

\usepackage[version=3]{mhchem} %
\usepackage{booktabs}          %
\usepackage{graphicx}
\usepackage{xcolor}
\usepackage{natbib}
\usepackage{hyperref}

\SectionNumbersOn

\author{Philippe Schwaller}
\affiliation{IBM Research -- Europe, Säumerstrasse 4, 8803 Rüschlikon, Switzerland}
\alsoaffiliation{Department of Chemistry and Biochemistry, University of Bern, Freiestrasse 3, 3012 Bern, Switzerland}
\email{phs@zurich.ibm.com}
\author{Daniel Probst}
\affiliation{Department of Chemistry and Biochemistry, University of Bern, Freiestrasse 3, 3012 Bern, Switzerland}
\author{Alain C. Vaucher}
\affiliation{IBM Research -- Europe, Säumerstrasse 4, 8803 Rüschlikon, Switzerland}
\author{Vishnu H. Nair}
\affiliation{IBM Research -- Europe, Säumerstrasse 4, 8803 Rüschlikon, Switzerland}
\author{David Kreutter}
\affiliation{Department of Chemistry and Biochemistry, University of Bern, Freiestrasse 3, 3012 Bern, Switzerland}
\author{Teodoro Laino}
\affiliation{IBM Research -- Europe, Säumerstrasse 4, 8803 Rüschlikon, Switzerland}
\author{Jean-Louis Reymond}
\affiliation{Department of Chemistry and Biochemistry, University of Bern, Freiestrasse 3, 3012 Bern, Switzerland}

\title{Mapping the Space of Chemical Reactions Using Attention-Based Neural Networks}

\begin{document}

\begin{abstract}
\textbf{}  Organic reactions are usually assigned to classes containing reactions with similar reagents and mechanisms. Reaction classes facilitate the communication of complex concepts and efficient navigation through chemical reaction space.  However, the classification process is a tedious task. It requires the identification of the corresponding reaction class template via annotation of the number of molecules in the reactions, the reaction center, and the distinction between reactants and reagents. This work shows that transformer-based models can infer reaction classes from non-annotated, simple text-based representations of chemical reactions. Our best model reaches a classification accuracy of 98.2\%. We also show that the learned representations can be used as reaction fingerprints that capture fine-grained differences between reaction classes better than traditional reaction fingerprints. The insights into chemical reaction space enabled by our learned fingerprints are illustrated by an interactive reaction atlas providing visual clustering and similarity searching. 
\end{abstract}

In the last decade, computer-based systems \cite{grzybowski_wired_2009,coley2017computer, rxn} have become an important asset available to chemists. Deep learning methods stand out, not only for reaction prediction tasks \cite{coley2017prediction, schwaller2018found, Schwaller_CentrScie_2019}, but also for synthesis route planning \cite{segler_planning_2018,thakkar2020datasets, schwaller2019predicting} and synthesis procedures to action conversions \cite{vaucher2020automated}. 

\begin{figure}[ht!]
  \centering
   \includegraphics[width=\linewidth]{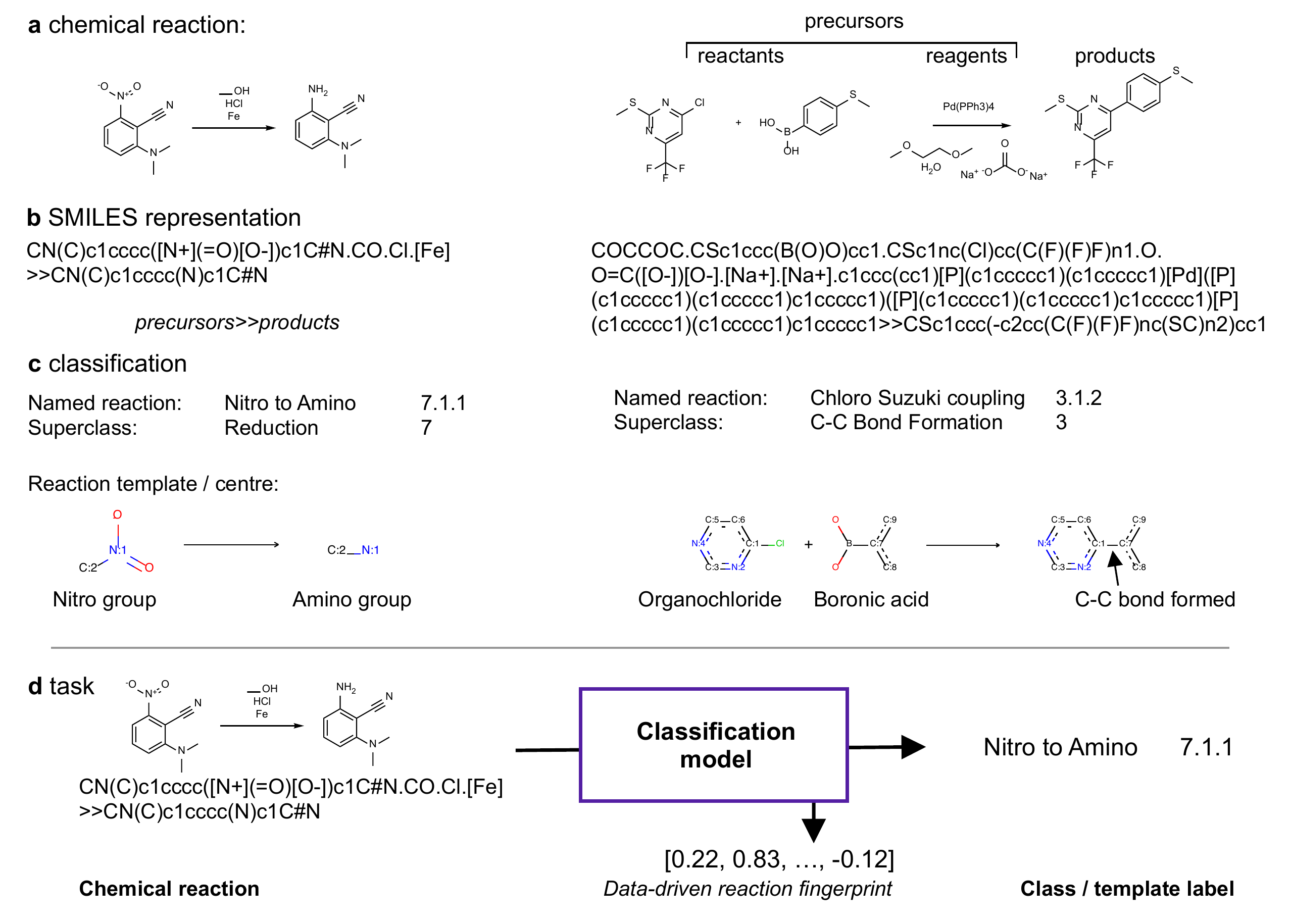}
  \caption{\textbf{Data representation and task.} Two examples of chemical reactions with associated  classification labels and reaction templates describing the transformation. The task is to predict the reaction class or template label from the chemical reaction. The encoded representation of the reaction can be used as data-driven reaction fingerprint.}
  \label{fig:intro}
\end{figure}

Among the few approaches, natural language processing (NLP) methods \cite{vaswani2017attention, devlin2019bert} applied to Simplified molecular-input line-entry system (SMILES) \cite{weininger_smiles_1988,weininger1989smiles} and other text-based representation of molecules and reactions are particularly effective in the chemical domain. Recently, Schwaller et al. \cite{Schwaller2020Unsupervised} demonstrated that neural networks are able to capture the atom rearrangements from precursors to products in chemical reactions without supervision. Figure \ref{fig:intro} a) shows examples of chemical reactions and the corresponding textual representation in b).

 The demand for robust algorithms to categorise chemical reactions is high. The knowledge of the class of a reaction has a great value for expert chemists, for example to assess the quality of the reaction prediction \cite{Toniato2020}. Chemists use reaction classes to navigate large databases of reactions and retrieve similar members of the same class to analyse and infer optimal reaction conditions. They also use reaction classes as an efficient way to communicate what a chemical reaction does and how it works in terms of atomic rearrangements. As seen in Figure \ref{fig:intro} c), reaction classes can be named after the reaction type referring to the changing structural features, such as ``Nitro to Amino''. Alternatively, they can be named after the persons who discovered the chemical reaction or refined an already known transformation, like the second example in Figure \ref{fig:intro} c). It is a chloro Suzuki coupling reaction named after Akari Suzuki, who received the Nobel prize in 2010 for his work on palladium-catalysed cross-coupling reactions \cite{miyaura1995palladium}. The current state-of-the-art in reaction classification is are commercially available tools \cite{NameRXN, kraut2013algorithm}, which classify reactions based on a library of expert-written rules. These tools typically make use of  SMIRKS \cite{smirks}, a language for describing transformations in the SMILES format \cite{weininger1988smiles, weininger1989smiles}. On the contrary, classifiers based on machine learning have the potential to increase the robustness to noise in the reaction equations and to avoid the need for an the explicit formulation of rules. 

Early work in the 90s used self-organising neural networks to map organic reactions and investigate similarities between them \cite{chen1996organic, chen1997knowledge, satoh1998classification}. More recently, Schneider et al.\cite{schneider2015development} developed a reaction classifier based on traditional reaction fingerprints. Molecular and reaction fingerprints are fixed-size vector encodings of discrete molecular structures and chemical reactions. The currently best performing fingerprint by Schneider et al.\cite{schneider2015development} combines a products-reactants difference fingerprint with molecular features calculated on the reagents and was tested on a limited set of 50 reaction classes. This difference fingerprint is currently one of the most frequently used hand-crafted ones. It has been successfully applied to reaction conditions predictions \cite{gao2018using}, where the reagents were not taken into account for the reaction description. Ghiandoni et al. \cite{ghiandoni2019development} introduced an alternative hierarchical classification scheme and random forest classifier for reaction classification. Their algorithm outputs a confidence score through conformal prediction. The fingerprints developed by Schneider et al. \cite{schneider2015development} and Ghiandoni et al. \cite{ghiandoni2019development} both require a reactants-reagents role separation \cite{schneider2016s}, which is often ambiguous and thus limits their applicability. 

Traditionally, reaction fingerprints were hand-crafted using the reaction center or a combination of the reactant, reagent and product fingerprints. ChemAxon \cite{chemaxon}, for instance, provides eight types of such reaction fingerprints. Based on the differentiable molecular fingerprint by Duvenaud et al. \cite{duvenaud2015convolutional}, the first example of a learned reaction fingerprint was presented by Wei et al. \cite{wei2016neural} and used to predict chemical reactions. Unfortunately, their fingerprint was restricted to a fixed reaction scheme consisting of two reactants and one reagent, and hence, only working for reactions conform with that scheme. Similarly, the multiple fingerprint features by Sandfort et al. \cite{sandfort2020structure} are made by concatenating multiple fingerprints for a fixed number of molecules.

In the first part of our work, we predict chemical reaction classes using attention-based neural networks from the family of transformers \cite{vaswani2017attention, devlin2019bert}. Our deep learning models do not rely on the formulation of specific rules that require every reaction to be properly atom-mapped. Instead, they learn the atomic motifs that differentiate reactions from differentclasses from raw reaction SMILES without reactant-reagent role annotations (Figure 1d). The transformer-based sequence-2-sequence (seq-2-seq) model \cite{vaswani2017attention} matched the ground-truth classification with an accuracy of 95.2\% and the Bidirectional Encoder Representations from Transformers (BERT) classifier\cite{devlin2019bert} with 98.2\%. We analyse the encoder-decoder attention of the seq-2-seq
model and the self-attention of the BERT model. Hereby we observe that atoms involved in the reaction center, as well as reagents specific to the reaction class, have larger attention weights. 

In the second part, we demonstrate that the representations learned by the BERT models, unsupervised and supervised, can be used as reaction fingerprints. The reaction fingerprints we introduce are independent of the number of molecules involved in a reaction. The BERT models trained on chemical reactions can convert any reaction SMILES into a vector without requiring atom-mapping or a reactant-reagent separation. Therefore our reaction fingerprints are universally applicable to any reaction database. Based on those reaction fingerprints and TMAP \cite{probst2019visualization}, a method to visualise high-dimensional spaces as tree-like graphs, we were able to map the chemical reaction space and show in our reaction atlases nearly perfect clustering according to the reaction classes. Moreover, our fingerprints enable chemists to efficiently search chemical reaction space and retrieve metadata of similar reactions. The metadata could, for instance, contain typical conditions, synthesis procedures, and reaction yields.

On an imbalanced data set, our fingerprints and classifiers reach an overall classification accuracy of more than 98\%, compared to 41 \% when using a traditional reaction fingerprint. The ability to accurately classify chemical reactions and represent them as fingerprints, enhances the accessibility of reaction by machines and humans alike. Hence, our work has the potential to unlock new insights in the field of organic synthesis. In recent studies, our models were used to predict experimentally measured activation energies\cite{Jorner2020} and reaction yields\cite{Schwaller2020}.

\section*{Results and Discussion}
\subsection*{Reaction classification}
\subsubsection*{Classification results}
We used a labeled set of chemical reactions as ground truth to train two transformer-based deep learning models as architecture \cite{vaswani2017attention,devlin2019bert}. The first one is an encoder-decoder transformer as introduced by Vaswani et al. \cite{vaswani2017attention} for sequence-to-sequence (seq-2-seq) tasks in neural machine translation. The second one is an encoder-only transformer called BERT introduced by Devlin et al. \cite{devlin2019bert}. The latter model with a classification head on top is typically used in NLP for single sentence classification tasks \cite{socher2013recursive,  warstadt2019neural}. A visualisation of such a BERT classifier is shown in Figure \ref{fig:model}.

The ground truth data is composed of chemical transformations represented in text format as SMILES. Their labeling (classification) was taken from the strongly imbalanced Pistachio data set \cite{Pistachio2017}, which uses NameRXN for the reaction classification \cite{NameRXN}. In an additional experiment, we use reaction template labels derived from open-source data, which we will refer to as USPTO 1k TPL. We analysed the classification performance of our models on the test sets, which contained 132k reactions from 792 different classes in Pistachio, and 45k reactions from 1000 template classes in USPTO 1k TPL.
A summary of the results can be found in Table \ref{tab:results}.
On the Pistachio test set, the transformer encoder-decoder model (enc2-dec1) matched the ground truth classification with an accuracy of 95.2\%. The reaction BERT classifier predicted the correct name reaction with an accuracy of 98.2\%, therefore achieving significantly better results than with the seq-2-seq approach. As a comparison to previous work \cite{schneider2015development}, we computed the transformation fingerprint AP3 (folded) +  featureFP on the Pistachio data and used a 5-NearestNeighbour (5-NN) classifier \cite{johnson2017billion} to classify the test set reactions. Even though we separated the reactants and reagents using RDKit \cite{greg_landrum_2019_3366468}, the classifier only achieved an overall accuracy of 41.0\%. The traditional fingerprint was not able to represent the fine-grained differences between the reaction classes. The ``Unrecognised'', ``Carboxylic acid + amine condensation'', ``Amide Schotten-Baumann'' and ``N-Boc deprotection'' classes contained the most false positives. 

\begin{table}[!ht]
\center
\caption{\textbf{Classification results.} The lower the confusion entropy of a confusion matrix (CEN) and the higher the Matthews correlation (MCC) coefficient the better. The traditional fingerprint is an AP3 256 (folded) + agents features developed by Schneider et al. \cite{schneider2015development}.}
    \begin{tabular}{l  r r r }
        \toprule  
\textbf{Pistachio} & Accuracy & CEN & MCC\\ \midrule
Traditional fp\cite{schneider2015development} + 5-NN classifier & 0.410 & 0.365 & 0.305\\
Transformer enc2-dec1 & 0.952 & 0.039 & 0.946\\
BERT classifier & \textbf{0.982} & \textbf{0.014} & \textbf{0.980}\\ 
\emph{rxnfp (pretrained)} + 5-NN classifier & 0.819 & 0.121 & 0.797\\
\emph{rxnfp} + 5-NN classifier & \textbf{0.989} & \textbf{0.010} & \textbf{0.988}\\ \midrule
\textbf{USPTO 1k TPL} & Accuracy & CEN & MCC\\ \midrule
Traditional fp\cite{schneider2015development} + 5-NN classifier &0.295 & 0.424 & 0.292\\
BERT classifier & \textbf{0.989} & \textbf{0.006} & \textbf{0.989}\\ 
\emph{rxnfp (pretrained)} + 5-NN classifier & 0.340 & 0.392 & 0.337\\
\emph{rxnfp} + 5-NN classifier & \textbf{0.989} & \textbf{0.006} & \textbf{0.989}\\
         \bottomrule
    \end{tabular}
    \label{tab:results}
\end{table}

In contrast, our BERT classifier without reactant-reagent separation was the best performing model, when looking at the confusion entropy of a confusion matrix (CEN) \cite{wei2010novel} and overall Matthews correlation coefficient (MCC) \cite{matthews1975comparison,gorodkin2004comparing}.

To show that the inferior performance of the traditional reaction fingerprint did not stem from the choice of the 5-NN classifier, we took the embeddings of the pretrained (\emph{rxnfp (pretrained})) and finetuned BERT (\emph{rxnfp}) as inputs for the 5-NN classifier. We then classified the test set reactions and computed the scores. As expected, the results for \emph{rxnfp}, which corresponds to the input of the classifier layer in the BERT classifier, perfectly matched the scores of the BERT classifier. 

The mismatches in the Pistachio test set are mainly related to ``Unrecognised'' reactions. When analysing the individual errors, we observed that our models were able to predict the correct reaction class for reactions that had a slight change in the representation between precursors and product (e.g.\ different tautomers). Such examples were not matched by the brittle rules that generated the ground-truth classes. Hence, they were labeled as ``Unrecognised'' reactions. Our models show very high robustness against errors in the SMILES representation. In the supplementary information, we report cases where, despite an error in the molecular representation, our model was able to correctly classify the reaction that was originally described by chemists in the patent procedure text.

On the USPTO 1k TPL test set, the traditional and pretrained fingerprint performed worse than on the Pistachio data set. However, the BERT classifier as well as the embeddings of the BERT classifier with the 5-NN classifier matched the performance they had on the Pistachio data set with an accuracy of 98.9\%.

An elaborate description of both types of reaction fingerprints is presented in the section on data-driven reaction fingerprints below. A comparison of our data-driven approach to traditional fingerprints on a balanced data set of 50k reactions can be found in the supplementary information. Even when using as little as 10k training reactions from 50 different classes the fine-tuned embeddings are able to outperform traditional fingerprints by increasing precision, recall and F1-score from 0.97 to 0.99.

\subsubsection*{Visualisation of Attention Weights}
\begin{figure}[ht!]
  \centering
   \includegraphics[width=\linewidth]{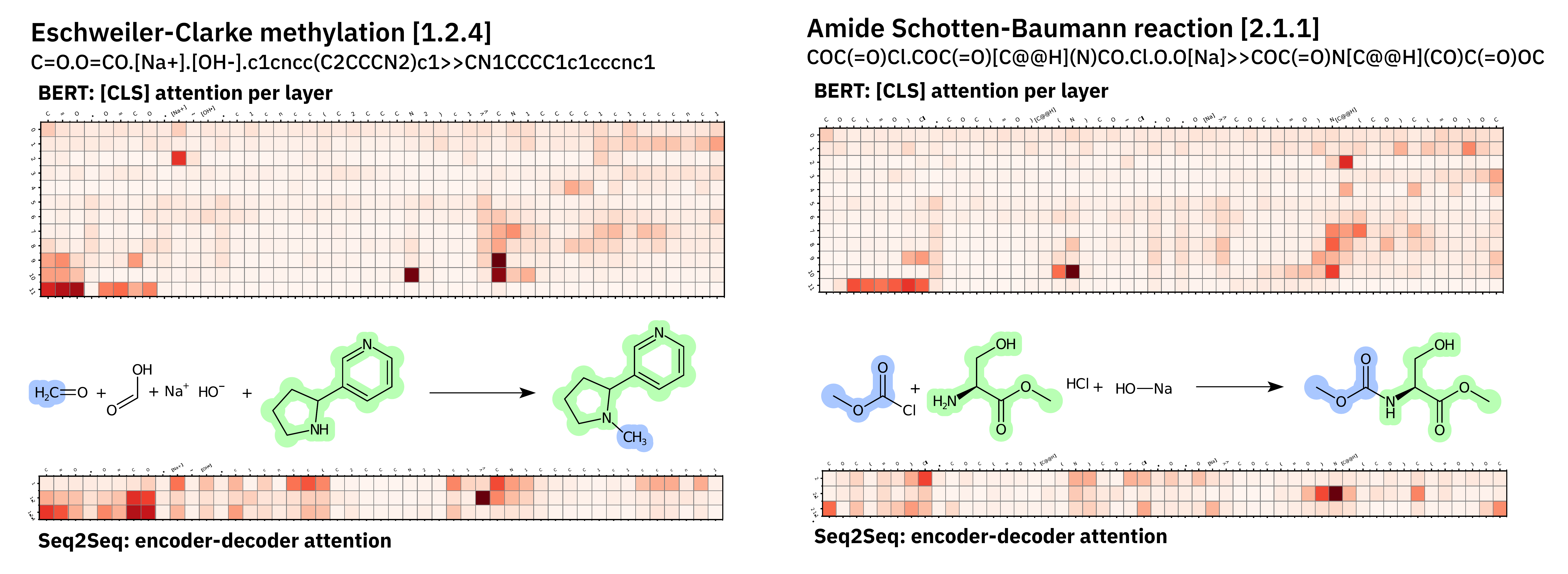}
  \caption{\textbf{Attention weights interpretation.} Layer-wise [CLS] token attention for the BERT classifier and 
  encoder-decoder attention for the enc2-dec1 transformer model. The horizontal axis contains the SMILES tokens of the input reaction. The darker the token the more attention a specific token had received in that particular layer or output step. The colouring on the reaction depictions created with CDK depict \cite{willighagen2017chemistry} shows the mapping from precursors to product in the ground truth.}
  \label{fig:attn}
\end{figure}
Figure \ref{fig:attn} shows the layer-wise [CLS] token attention of the BERT classifier (above the reaction) and the encoder-decoder attention of the seq-2-seq model (below the reaction) for two different chemical transformations. We observed that the larger weights were associated with the atoms that are part of the reaction center or precursors specific to the reaction class. Just like a human expects to see a certain group of atoms based on the classification, for the seq-2-seq model, the decoder learned to focus on the atoms involved in the rearrangement to classify reactions. For the BERT classifier, the initial layers had weak attention on all reaction tokens. The middle layers tended to attend either the product or the precursors. The  last layers focused on the reaction center and the precursors that are important for the classification.

\subsection*{Mapping Chemical Reaction Space}
\label{sec:rxnpf}
\subsubsection*{Data-driven Reaction Fingerprints}
 Molecular fingerprints are widely used to screen molecules with similar properties or map chemical space \cite{capecchi2020one}. 
Our reaction BERT models does not only perform best on the classification task but also allows chemists to generate vectorial representations of chemical reactions. Here we introduce reaction fingerprints based on the embeddings computed by BERT \cite{devlin2019bert} models. They can be applied to any reaction data set, as they do not require a reactant-reagent split or a fixed number of precursors.
During the pretraining of the BERT model, individual tokens in the reaction SMILES are masked and then predicted by the model. As the prepended [CLS] token is never masked, the model is always able to attend the representation of this token to recover the masked tokens. The intuition is that the model uses the [CLS] token to embed a global description of the reaction. Before the fine-tuning, the [CLS] token embeddings are learned purely by self-supervision. We refer to this fingerprint as \emph{rxnfp (pretrained)}. For the supervised fine-tuning, the embeddings of the [CLS] token are then taken as input for a one layer classification head and further refined. We refer to the fingerprint fine-tuned on the Pistachio training set as \emph{rxnfp}.
In our case, the [CLS] token embedding is a vector of size 256, corresponding to the hidden size of the BERT model. 
During the supervised classification task, the model has to focus on the reaction center and certain precursors that are specific to the individual name reactions. For instance, the Eschweiler-Clarke methylation (1.2.4) is a methylation reaction that can be distinguished from other methylation reactions as its precursors contain formaldehyde and formic acid (see Figure \ref{fig:attn}). Another example are Suzuki-type coupling reactions, where the ``-type'' suffix means that the metal catalyst is missing but the described reaction would otherwise correspond to a Suzuki coupling reaction.

\subsubsection*{Reaction Atlases}
\begin{figure}[ht!]
  \centering
   \includegraphics[width=\linewidth]{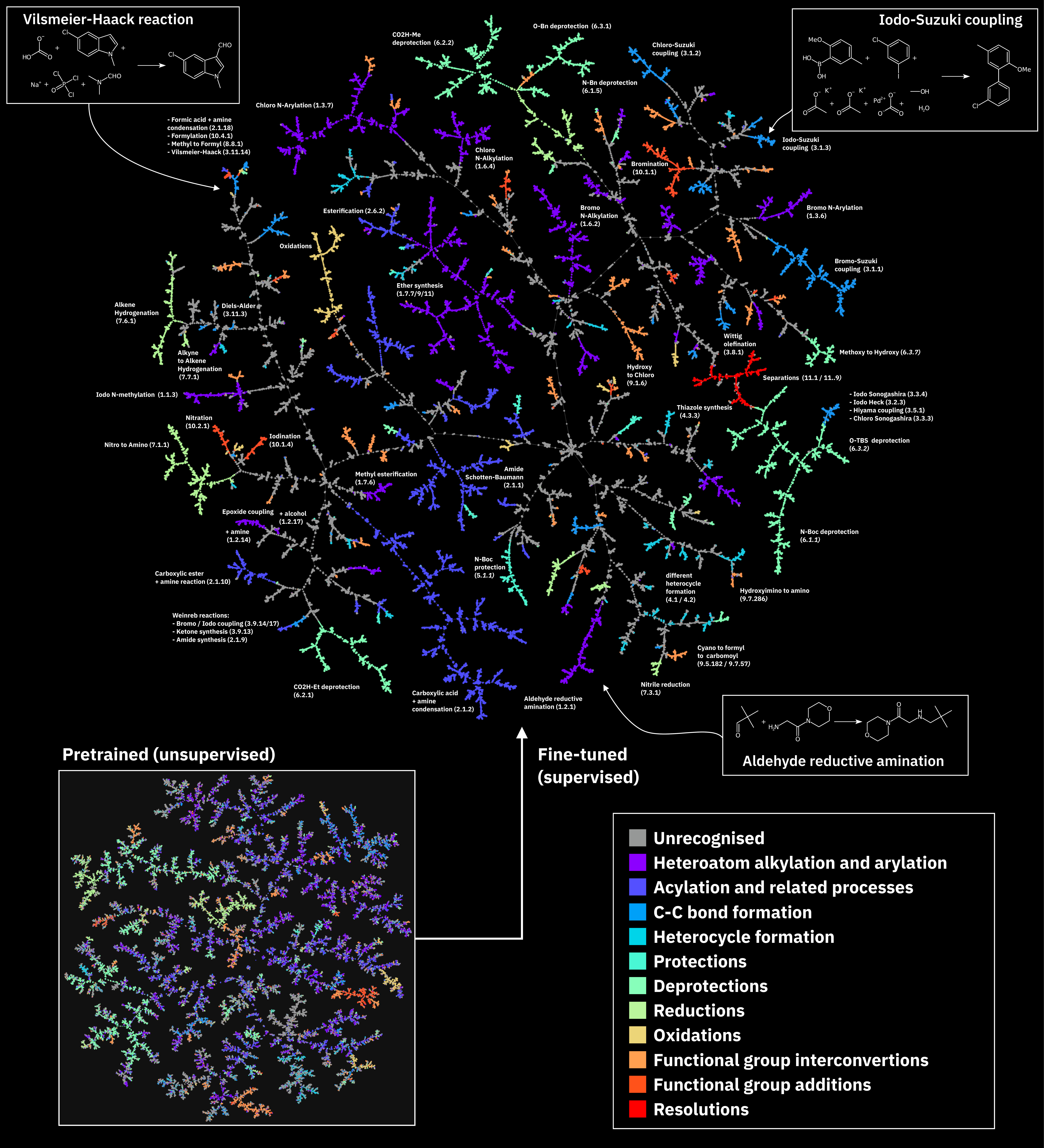}
  \caption{\textbf{Reaction atlases.} Top: Annotated reaction atlas created from \emph{rxnfp}. Bottom: reaction atlas made from \emph{rxnfp (pretrained)}. The different fingerprints of the test set reactions are visualised using a TMAP algorithm \cite{probst2019visualization} and the Faerun visualiation library \cite{probst2017fun}. The fingerprints were minhashed using a weighted hashing scheme to make them compatible with the LSH forest.}
  \label{fig:ann_atlas}
\end{figure}

In Figure \ref{fig:ann_atlas}, we show an annotated version of a reaction atlas created by using the embeddings of a BERT classifier fine-tuned for three epochs. The colours correspond to the 12 superclasses found in the data set. The individual classes are almost perfectly clustered. It is worth noting that the sub-trees in the TMAP closely group related reaction classes.  For instance, in the upper left, one sub-tree contains all ``Formylation''-related reactions, Weinreb reactions are clustered in a branch in the lower left and Suzuki-type reactions share the same branch as the corresponding Suzuki reactions. The unannotated reaction atlas was created using the fingerprints computed from a pretrained reaction BERT model without classification fine-tuning. Even after applying a purely unsupervised masked language modeling training, the model was already able to extract features relevant for reaction classification and some clustering can be observed in the figure.

An interactive reaction TMAP \cite{probst2019visualization}, visualising the public Schneider 50k \cite{schneider2015development} data set by  using the \emph{rxnfp (10k)} embeddings and highlighting different precursor and product properties, can be found on \url{https://rxn4chemistry.github.io/rxnfp/tmaps/tmap_ft_10k.html}.

\subsubsection*{Reaction search}
One of the primary use cases for reaction fingerprints is the search for similar reactions in a database.  An atom-mapping independent reaction fingerprint is extremely powerful, as it unlocks the possibility of reaction retrieval without the need of knowing the reaction center. For instance, when a black box model like a forward reaction prediction model \cite{Schwaller_CentrScie_2019} or a retrosynthesis model \cite{schwaller2019predicting} predicts a reaction, the most similar reactions from the training set of those models could be retrieved. Such a retrieval of similar reactions could not only increase the explainability of deep learning models. It would also allow chemists to access the metadata (including yield and reaction conditions) of the closest reactions, if this information is available.

\begin{figure}[ht!]
  \centering
   \includegraphics[width=\linewidth]{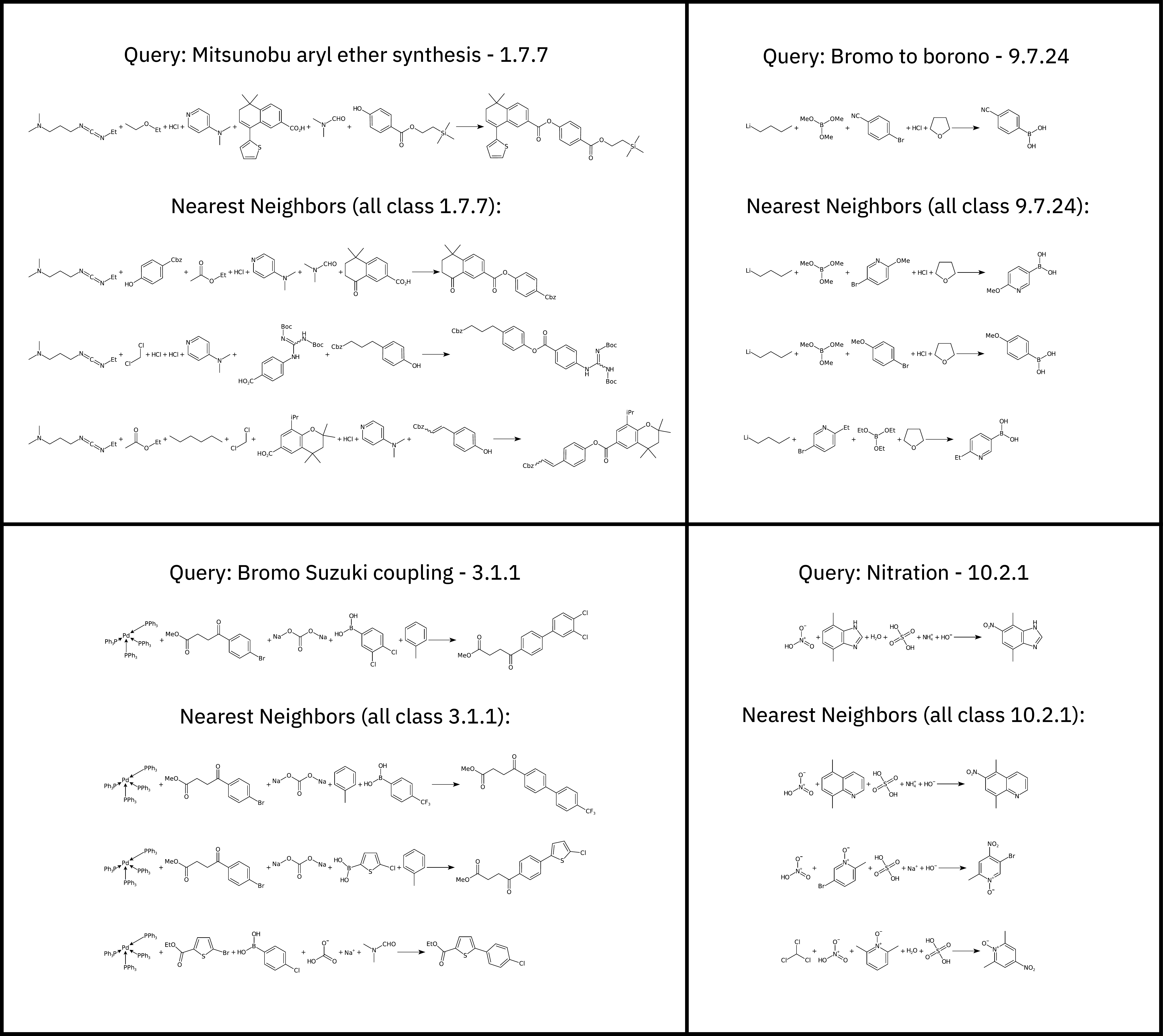}
  \caption{\textbf{Nearest-neighbour queries.} Four examples of reaction SMILES queries and the three nearest neighbours retrieved from the LSHForest \cite{probst2019visualization} of the training set containing 2.4M reactions. All the retrieved reactions belong to the same reaction class as the query reaction and show similar precursors.}
  \label{fig:nn_queries}
\end{figure}

In Figure \ref{fig:nn_queries} the three approximate nearest neighbours of the BERT classifier fingerprint can be found for four test set reactions from four distinct reaction classes. The nearest neighbours searches on the training set containing 2.4M reactions were performed within milliseconds using unoptimised python code on a MacBook Pro (Processor: 2.7 GHz Intel Core i7, Memory: 16 GB 2133 MHz LPDD). They were based on the LSH forest from the TMAP module developed by Probst and Reymond \cite{probst2019visualization} In all searches, the nearest neighbours corresponded to the same class as the query reaction. The similarities between the query reaction and the retrieved nearest neighbours were clearly visible even for non-experts. The reactions share similar, if not the same precursors, and the products show similar features. One of the great advantages of this reaction search method is that it only requires a reaction SMILES as input.

To investigate the robustness of our BERT classifier embeddings we removed three classes from the fine-tuning training set (Number of removed reaction classes: {`1.6.4 - Chloro N-alkylation': 24109, `3.9.17 - Weinreb Iodo coupling': 225, `9.7.73 - Hydroxy to azido': 1526}) and fine-tuned another BERT classifier. After 5 epochs, we generated the embeddings for the test set reactions from  the three removed classes. For the ``Chloro N-alkylation'' and the ``Hydroxy to azido'' class the most common prediction was ``Unrecognised''. All the predictions of the BERT model trained without the removed classes for the ``Weinreb Iodo coupling'' were ``Weinreb bromo coupling'' that differs just by the type of the reacting halogen atom. Another interesting experiment is the retrieval of nearest neighbours from the original training set for the embeddings generated by the BERT model trained without the removed classes. For 1078 out of 1370 ``Chloro N-alkylation'' reactions in the test set, the nearest neighbour in the initial training set (including all the reaction classes) was a ``Chloro N-alkylation'' reaction. For the 10 ``Weinreb Iodo coupling'' reactions, the nearest neighbours in the original training set were four ``Weinreb Bromo coupling'' and other four ``Bromo Grignard + nitrile ketone synthesis'' reactions, which are both closely related reaction types. There was no clearly dominating reaction class in the nearest neighbours with 44 out of 76 reactions being ``Unrecognised''.

\section*{Conclusion}
In this work, we focused on the data-driven classification of chemical reactions with natural language processing methods and on the use of their embedded information to design reaction fingerprints.
Our transformer-based models were able to learn the classification schemes using a broad set of chemical reactions as ground-truth, labeled by a  commercially available reaction classification tool. With the BERT classifier, we match the rule-based classification with an accuracy of 98.2\.\%, compared to 41\% for a traditional fingerprint plus 5-nearest neighbours classifier. 
Our models are able to learn the atomic environment characteristics of each class and provide a rationale that is easily interpretable by chemists. Understanding the reasoning behind each classification by using the attention weights may help the end-user chemists with the adoption process of these technologies. We showed that the representations learned by our BERT models can be used as reaction fingerprints. Those data-driven reaction fingerprints unlock the possibility of mapping the reaction space without knowing the reaction centers or the reactant-reagent split. They also enable efficient nearest neighbour searches on reaction data sets containing millions of reactions. Moreover, our fingerprints were recently used to estimate experimentally measured activation energies \cite{Jorner2020} and fine-tuned to predict chemical reaction yields \cite{Schwaller2020}.

\section*{Methods}

\subsection*{Data}
The data consisted of 2.6M reactions extracted from the Pistachio database \cite{Pistachio2017} (version 191118), where we removed duplicates and filtered invalid reactions using RDKit \cite{greg_landrum_2019_3366468}. The data set was split into train, validation and test sets (90\%\,/\,5\%\,/\,5\%), with reactions with identical products kept in the same set. The reaction data in Pistachio was classified using NameRXN \cite{NameRXN}, a rule-based software that classifies roughly 1000 different name reactions. The classification is organised into superclasses \cite{carey2006analysis}, reaction categories and name reactions according to the RXNO ontology \cite{rxno}. For more detail on name reactions and their categories, we refer the reader to the work of Schneider et al.\cite{schneider2016big}.  As common in practice, we represent the chemical reactions with reaction SMILES \cite{weininger1988smiles, weininger1989smiles}. We tokenise the reaction SMILES as in Schwaller et al.\cite{Schwaller_CentrScie_2019} without enforcing any distinction between reactants and reagents. Therefore, our method is universally applicable, including those reactions where the reactant-reagent distinction is subtle \cite{schneider2016s}. To compare with previous work and ensure reproducibility, we used the reaction data set published by Schneider et al. \cite{schneider2015development} with 50k reactions belonging to 50 different reaction classes. We also introduced an open-source reaction classification data set, which we named USPTO 1k TPL, derived from the USPTO data base by Lowe \cite{Lowe2017}. It consists of 445k reactions divided into 1000 template labels. The data set was randomly split into 90\% for training and validation and 10\% for testing. The labels were obtained by atom-mapping the USPTO data set with RXNMapper \cite{Schwaller2020Unsupervised}. Subsequently, the template extraction workflow by Thakkar et al. \cite{thakkar2020datasets,coley2019rdchiral} was applied and finally, selecting reactions belonging to the 1000 most frequent template hashes. Those template hashes were used as class labels. Similarly to the Pistachio data set, USPTO 1k TPL is strongly imbalanced.

\subsection*{Models}

We trained two different types of deep learning models inspired by recent progress in Natural Language Processing. The first model is an autoregressive encoder-decoder transformer model \cite{vaswani2017attention}. We constructed the model with 2 encoder layers and 1 decoder layer. For the prediction target, we split the class prediction into superclass, category and name reaction prediction. This means, for example, that the target string for the name reaction ``1.2.3'' would be ``1 1.2 1.2.3''. As the source and target are dissimilar, we did not share encoder and decoder embeddings. We used the same remaining hyperparameter as were used for the training of the Molecular Transformer \cite{Schwaller_CentrScie_2019, opennmt}, which is state-of-the-art in chemical reaction prediction. 

\begin{figure}[ht!]
  \centering
   \includegraphics[width=\linewidth]{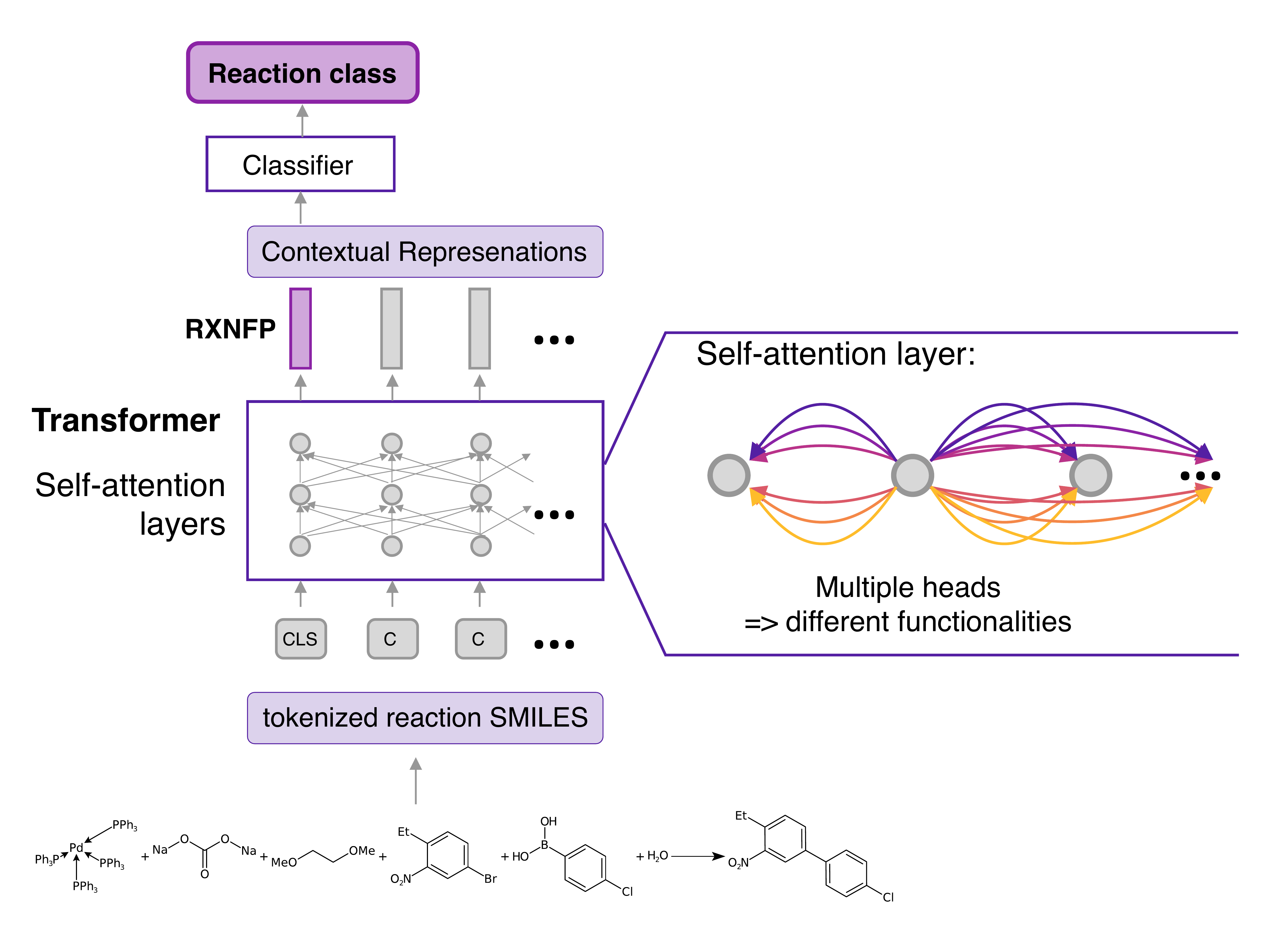}
  \caption{\textbf{BERT reaction classification model} The figure illustrates a BERT model with stacks of self-attention layers. All self-attention layers consist of multiple attention heads. Using a classifier head the model was applied to a chemical reaction classification task. The encoding of the [CLS] token can also be used as reaction fingerprint (rxnfp).}
  \label{fig:model}
\end{figure}

One of the major recent advancement in natural language processing is BERT \cite{devlin2019bert}, which compared to the seq-2-seq architecture only consists of a transformer encoder with specific heads that can be fine-tuned for different tasks such as multi-class prediction. The model is visualised in Figure \ref{fig:model}. We pretrained a BERT model using masked language modeling loss on the chemical reactions. The task of the model in masked language modeling consists of predicting individual tokens of the input sequence that have been masked with a probability of 0.15. Same as in the BERT training, a special class token [CLS] was prepended to the tokenised reaction SMILES. The [CLS] token was never masked during this self-supervised training. In contrast to the original BERT pretraining \cite{devlin2019bert}, we did not use the next sentence prediction task. We then fine-tuned the pretrained model with a classifier head on the name reaction classes. The embeddings of the [CLS] token were taken as input for the classifier head. 
Compared to the hyperparameters of the BERT-Base model in Ref.~\citenum{bertgithub}, we decreased the hidden size to 256, the intermediate size to 512, and the number of attention heads to 4. For the pretraining, we set 820k steps with a learning rate of 1e-4 and a maximum sequence length of 512, the rest of the parameters were kept as suggested in Ref.~\citenum{bertgithub}. For the classification fine-tuning, we only changed the learning rate to 2e-5, kept the maximum sequence length of 512 and fine-tuned for 5 epochs. After training, we converted the models to PyTorch \cite{NEURIPS2019_9015} models, which matched the Huggingface \cite{Wolf2019HuggingFacesTS} interface, as it facilitated further analysis. 

\subsection*{k-Nearest Neighbour Classifier}
The k-nearest neighbour classifier used to assess the quality of the proposed reaction representations is based on the FAISS framework developed by Facebook research \cite{johnson2017billion}. As FAISS provides an efficient implementation of brute-force k-nearest neighbour searches that can be applied on relatively large data sets.  Possible biases introduced through approximation methods were therefore avoided. The number of nearest neighbours $k=5$ and the Euclidean metric (L2) are chosen for all tests. The predicted class of the query was assumed to be the the one that is represented within most often the result set. Ties were broken using the distance between the query and one or more neighbours.

\subsection*{TMAP}
TMAP \cite{probst2019visualization} is a dimensionality reduction algorithm capable of handling millions of data points. The advantage of TMAP compared to other dimensionality reduction algorithms is the 2-dimensional tree-like output, which preserves both local and global structures, with a focus of local structure. The algorithm consists of four steps: 1) LSH Forest-based indexing, 2) k-nearest neighbour graph generation, 3) minimum spanning tree calculation using Kurskal's algorithm and 4) creating the tree-like layout. The resulting layout is then displayed using the interactive data visualisation framework Faerun \cite{probst2017fun}.

TMAP \cite{probst2019visualization} and Faerun \cite{probst2017fun} were originally developed to visualise large molecular data sets, but have been shown to be applicable to a wide range of other data. Here, we extended the framework with a customised version of SmilesDrawer \cite{probst2018smilesdrawer} that has been extended to allow for the display of chemical reactions.

\subsection*{Evaluation metrics}

To compare the results on the imbalanced classification test set, we used the confusion entropy of the confusion matrix (CEN) \cite{wei2010novel} calculated as follows,
\begin{equation*}
    P_{i,j}^{j}=\frac{Matrix(i,j)}{\sum_{k=1}^{|C|}\Big(Matrix(j,k)+Matrix(k,j)\Big)},
    \qquad
    P_{i,j}^{i}=\frac{Matrix(i,j)}{\sum_{k=1}^{|C|}\Big(Matrix(i,k)+Matrix(k,i)\Big)} 
\end{equation*}
\begin{equation*}
    CEN_j=-\sum_{k=1,k\neq j}^{|C|}\Bigg(P_{j,k}^jlog_{2(|C|-1)}\Big(P_{j,k}^j\Big)+P_{k,j}^jlog_{2(|C|-1)}\Big(P_{k,j}^j\Big)\Bigg)
\end{equation*}
\begin{equation*}
    P_j=\frac{\sum_{k=1}^{|C|}\Big(Matrix(j,k)+Matrix(k,j)\Big)}{2\sum_{k,l=1}^{|C|}Matrix(k,l)} 
\end{equation*}
\begin{equation*}
    CEN=\sum_{j=1}^{|C|}P_jCEN_j
\end{equation*}
where Matrix is the confusion matrix, and the overall Matthews Correlation Coefficient (MCC)\cite{matthews1975comparison, gorodkin2004comparing} is,

\begin{equation*}
    cov(X,Y)=\sum_{i,j,k=1}^{|C|}\Big(Matrix(i,i)Matrix(k,j)-Matrix(j,i)Matrix(i,k)\Big)
\end{equation*}
\begin{equation*}
    cov(X,X) = \sum_{i=1}^{|C|}\Bigg[\Big(\sum_{j=1}^{|C|}Matrix(j,i)\Big)\Big(\sum_{k,l=1,k\neq i}^{|C|}Matrix(l,k)\Big)\Bigg]
\end{equation*}
\begin{equation*}
    cov(Y,Y) = \sum_{i=1}^{|C|}\Bigg[\Big(\sum_{j=1}^{|C|}Matrix(i,j)\Big)\Big(\sum_{k,l=1,k\neq i}^{|C|}Matrix(k,l)\Big)\Bigg]
\end{equation*}

\begin{equation*}
    MCC=\frac{cov(X,Y)}{\sqrt{cov(X,X)\times cov(Y,Y)}}.
\end{equation*}

Both are recommended metrics for imbalanced multi-class classification problems. We computed the scores using PyCM \cite{haghighi2018pycm}. For the comparison on the balanced data set, we used the average recall, precision and F1 score, as those metrics were used by Schneider et al. \cite{schneider2015development}. The recall, precision and F1 score values for the individual classes are shown in the supplementary material.  

\section*{Data availability}
The Schneider 50k data set is publicly available \cite{schneider2015development}. We provide a new reaction data set (USPTO 1k TPL), derived from the work of Lowe \cite{Lowe2017}, containing the 1000 most common reaction templates as classes. It can be accessed through \url{https://rxn4chemistry.github.io/rxnfp}. The commercial Pistachio (version 191118) data set can be obtained from NextMove Software \cite{Pistachio2017}. Pistachio relies on Leadmine \cite{lowe2015leadmine} to text-mine patent data. The data set comes with reaction classes assigned using NameRXN (\url{https://www.nextmovesoftware.com/namerxn.html}).

\section*{Code availability}
The rxnfp code and the experiments on the public data sets, as well as an interative TMAP, can be found on \url{https://rxn4chemistry.github.io/rxnfp}\cite{zenodo}.

\section*{Correspondence}
The corresponding author is Philippe Schwaller.

\section*{Acknowledgement}
DP and JLR acknowledge financial support by the Swiss National Science Foundation (NCCR TransCure). We thank Linda Rudin for the careful proof-reading of our manuscript. 

\section*{Contributions}

PS and AV conceived the initial idea for the project. PS, DP, AV, VH trained models, performed the classification experiments and analysed the results. PS investigated the reaction fingerprints and wrote the code base. PS, DP and DK worked on the reaction atlases. The project was supervised by TL and JLR. All authors took part in discussions and contributed to the writing of the manuscript.

\section*{Competing interests}
The authors declare no competing interests.

\end{document}


\section{Reaction properties atlases}

\begin{figure}[ht!]
  \centering
   \includegraphics[width=\linewidth]{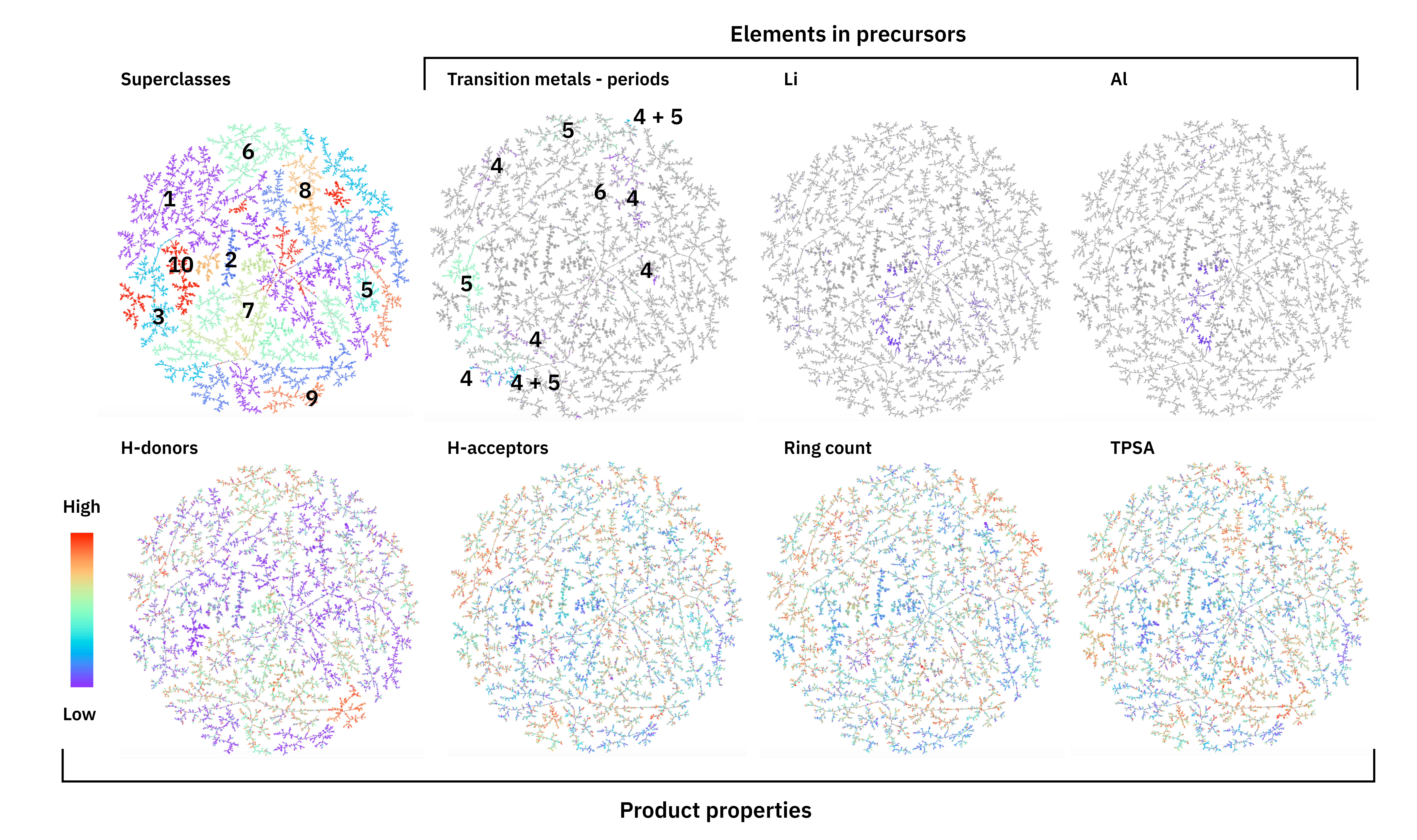}
  \caption{\textbf{Reaction properties} TMAP \cite{probst2019visualization} of the Schneider 50k set using the \emph{rxnfp (10k)} embeddings. The superclasses, as well as specific metallic elements in the precursors and product properties are highlighted in the different maps. An interactive version of this map is also available as a separate file.  }
  \label{fig:properties}
\end{figure}

Figure \ref{fig:properties} shows the chemical reaction found in the 50k set by \citet{schneider2015development} visualised with TMAP \cite{probst2019visualization} using the \emph{rxnfp (10k)}. The BERT model, which generated this reaction fingerprint was trained on the 10k training reactions. The reaction maps are made of the 10k training reactions plus 40k unseen reactions. The reactions corresponding to same reaction classes are well clustered together. We highlight reactions that contain specific elements in the precursors and observe that they found in the same branches of the map. Moreover, we visualize product properties and also observe defined clustering. 

\section{Analysis of Pistachio predictions}
We analysed the BERT classifier in more detail and compared it to the seq-2-seq transformer model. First, we identified different types of incorrect predictions by the transformer BERT classifier model, which are summarised in Table~\ref{tab:error_types}.
Most errors are related to the ``Unrecognised'' class of the RXNO ontology.
The most frequent error type is the prediction of a reaction class for a reaction classified as ``Unrecognised'' (47.9\% of all incorrect predictions), and the second most frequent error type is predicting ``Unrecognised'' when a class should be predicted (22.8\%). The third most frequent error is predicting the incorrect name reaction (third number of the class string, 17.5\%). The remaining errors are predicting an incorrect superclass (first number of the class string, 8.3\%) and predicting an incorrect category (second number of the class string, 3.5\%).

\begin{table}[!ht]
\center
\caption{\textbf{Incorrect predictions.} Types of incorrect predictions of the BERT model on the test set consisting of a total of 132213 reactions.}
\label{tab:error_types}
\begin{tabular}{lrr}
\toprule
                                                       & Count   & Percentage \\
\midrule
Correctly predicted& 129892& 98.24\% \\
Model predicts name reaction instead of ``Unrecognised''& 1111& 0.84\% \\
Model predicts ``Unrecognised'' instead of name reaction& 529& 0.40\% \\
Incorrect name rxn& 407& 0.31\% \\
Incorrect superclass& 193& 0.15\% \\
Incorrect category& 81& 0.06\% \\
\bottomrule
\end{tabular}
\end{table}

In Table~\ref{tab:worst_classes}, we show the reaction classes for which our model makes incorrect predictions most frequently.
Due to statistical sampling, we restricted this analysis to reactions with at least 20 occurrences in the test set.
For 12 out of 15 of these reaction classes, the most common error source is the failure to assign a reaction class, thus predicting ``Unrecognised''. Among the other most common failures, there is the ``Bouveault-Blanc reduction'', where an ester is reduced to a primary alcohol. Hence, it is very similar to the Ester to alcohol reduction class, with which it is most mistaken. The difference lies in the specific precursors used in the ``Bouveault-Blanc reduction'', such as sodium and ethanol or methanol.  The ``1,3-Dioxane synthesis'' reaction class has an overall accuracy of 88.9\%. However, there are some reactions mistaken for  ``Dioxolane synthesis'', for which the newly formed heterocycle in the product has an additional carbon atom.

\addtolength{\tabcolsep}{-3pt}
\begin{table}[!ht]
\center
\caption{\textbf{Detailed failure analysis.} Worst-predicted reaction classes with more than 20 occurrences in the test set for the BERT classifier.}
\label{tab:worst_classes}
\begin{tabular}{rlcrl}
\toprule
\multicolumn{2}{l}{Reaction class} & Accuracy [\%] & \multicolumn{2}{l}{Most frequent incorrectly predicted class} \\
\midrule
1.1.2& Menshutkin reaction& 62.1 & 0.0 & Unrecognised \\
3.9.41& Decarboxylative coupling& 72.1 & 0.0 & Unrecognised \\
9.7.140& Defluorination& 75.6 & 0.0 & Unrecognised \\
7.4.2& Bouveault-Blanc reduction& 76.4 & 7.4.1 & Ester to alcohol reduction \\
11.1& Chiral separation& 83.6 & 0.0 & Unrecognised \\
8.8.11& Hydroxylation& 83.7 & 0.0 & Unrecognised \\
4.3.11& Thiazoline synthesis& 85.7 & 0.0 & Unrecognised \\
3.9.12& Olefin metathesis& 85.8 & 0.0 & Unrecognised \\
2.5.5& Nitrile + amine reaction& 86.0 & 0.0 & Unrecognised \\
9.7.42& Chloro to fluoro& 86.4 & 0.0 & Unrecognised \\
10.4.2& Methylation& 88.9 & 0.0 & Unrecognised \\
4.2.39& 1,3-Dioxane synthesis& 88.9 & 4.2.20 & Dioxolane synthesis \\
4.1.53& 1,2,4-Triazole synthesis& 90.0 & 0.0 & Unrecognised \\
1.1.6& Chloro Menshutkin reaction& 90.6 & 0.0 & Unrecognised \\
5.1.2& N-Cbz protection& 90.9 & 2.1.1 & Amide Schotten-Baumann \\
\bottomrule
\end{tabular}
\end{table}
\addtolength{\tabcolsep}{3pt}

Although the large number of ``Unrecognised'' reactions in Pistachio makes an extensive analysis difficult, the inspection of a few dozen cases provides interesting insights. Part of the ``Unrecognised'' reactions should actually belong to a name reaction. The data-driven approach can be more robust than rule-based models and assign the correct reaction class. For example, in contrast to rule-based models, data-driven ones are often able to capture the reaction class despite changes in the tautomeric state between precursors and product. 
Another part of those ``Unrecognised'' reactions belongs to the category for which multiple transformations occur simultaneously. In this case, the reaction cannot be classified into a single name reaction, and our model predicts one of the corresponding reactions. Such examples can be found in deprotection reactions where more than one distinct functional group is removed. 
Another interesting aspect comes from molecules that are incorrectly parsed in Pistachio. If the SMILES string of a molecule involved in the reaction was incorrectly derived from the name, rule-based approaches fail to recognise the atomic rearrangements and thus to classify the reaction. For minor parsing errors, our model shows its potential, recognizing the correct transformation in several instances. 

The accuracy of the enc2-dec1 seq-2-seq model was 3\% worse than the one of the BERT classifier. When comparing the predictions of the two models, we observe that most of the differences are related to the ``Unrecognised'' class. 3511 out of 5108 reactions that were correctly predicted by the BERT classifier but not the seq-2-seq model belong to the ``Unrecognised'' class. Moreover, the three classes containing the most examples of reaction classes predicted correctly by the BERT classifier but not by the seq-2-seq model were ``Carboxylic acid + amine condensation'' (2.1.2), ``Methylation'' (10.4.2) and ``Williamson ether synthesis'' (1.7.9) reactions with 90, 61 and 37 examples respectively. In contrast, the seq-2-seq model was able to classify 474 reactions as ``Unrecognised'', which were classified as recognised name reactions by the BERT model. Besides the ``Unrecognised'' reactions, the three reaction types with the most examples that were correctly predicted by the seq-2-seq model but not by the BERT classifier were ``Bouveault-Blanc reduction'' (7.4.2), ``Ester to alcohol reduction'' (7.4.1) reactions with 33 and 15 examples respectively. The seq-2-seq seems to capture the subtle difference between the two distinct ``Ester to alcohol'' (7.4) classes better.

\section{Analysis of 50k set predictions}

\citet{schneider2015development} evaluated their reaction fingerprints by analysing how well it could classify chemical reactions using a logistic regression classifier \cite{scikit-learn}. For a given reaction input, they trained their classifier to predict 1 out of 50 named reaction classes using 200 training/validation and 800 testing examples per class. To be able to directly compare to the results of Ref.~\citenum{schneider2015development}, we investigated our learned fingerprints on their data sets, pretrained and fine-tuned on the same 10k training reactions resulting in \emph{rxnfp (10k)}. A summary where we report recall, precision and F-score averaged over the 50 classes can be found in Table \ref{tab:resultsfp}. While the \emph{rxnfp (pretrained)} does not suffice to match the performance of the handcrafted fingerprint on this balanced data set, \emph{rxnfp (10k)}, generated after fine-tuning the model on as little as the 10k reactions, is able to reach scores of 0.99 compared to 0.97 for the hand-crafted fingerprint.  

\begin{table}[!ht]
\center
\caption{Comparing fingerprints on the 50k reactions classification benchmark by \citet{schneider2015development} (50 classes, 1000 reactions per class, 200 for training/validation and 800 for testing) }
\begin{tabular}{@{}lllll@{}}
\toprule
Fingerprint & recall & precision & F-score &  \\
\midrule
AP3 256 (folded) \cite{schneider2015development}  & 0.97  & 0.97 & 0.97 & handcrafted,  \\
+ Agent features &  &  & & reactants-reagents separation \\ \midrule
\emph{rxnfp (pretrained)} & 0.90 & 0.90  & 0.90 & after pretraining\\
\emph{rxnfp (10k)} & 0.99 & 0.99 & 0.99 & fine-tuning on 10k reactions \\
 &  &  & & training set\cite{schneider2015development} \\ 
 \bottomrule
\end{tabular}
    \label{tab:resultsfp}
\end{table}

\begin{figure}[ht!]
  \centering
   \includegraphics[width=\linewidth]{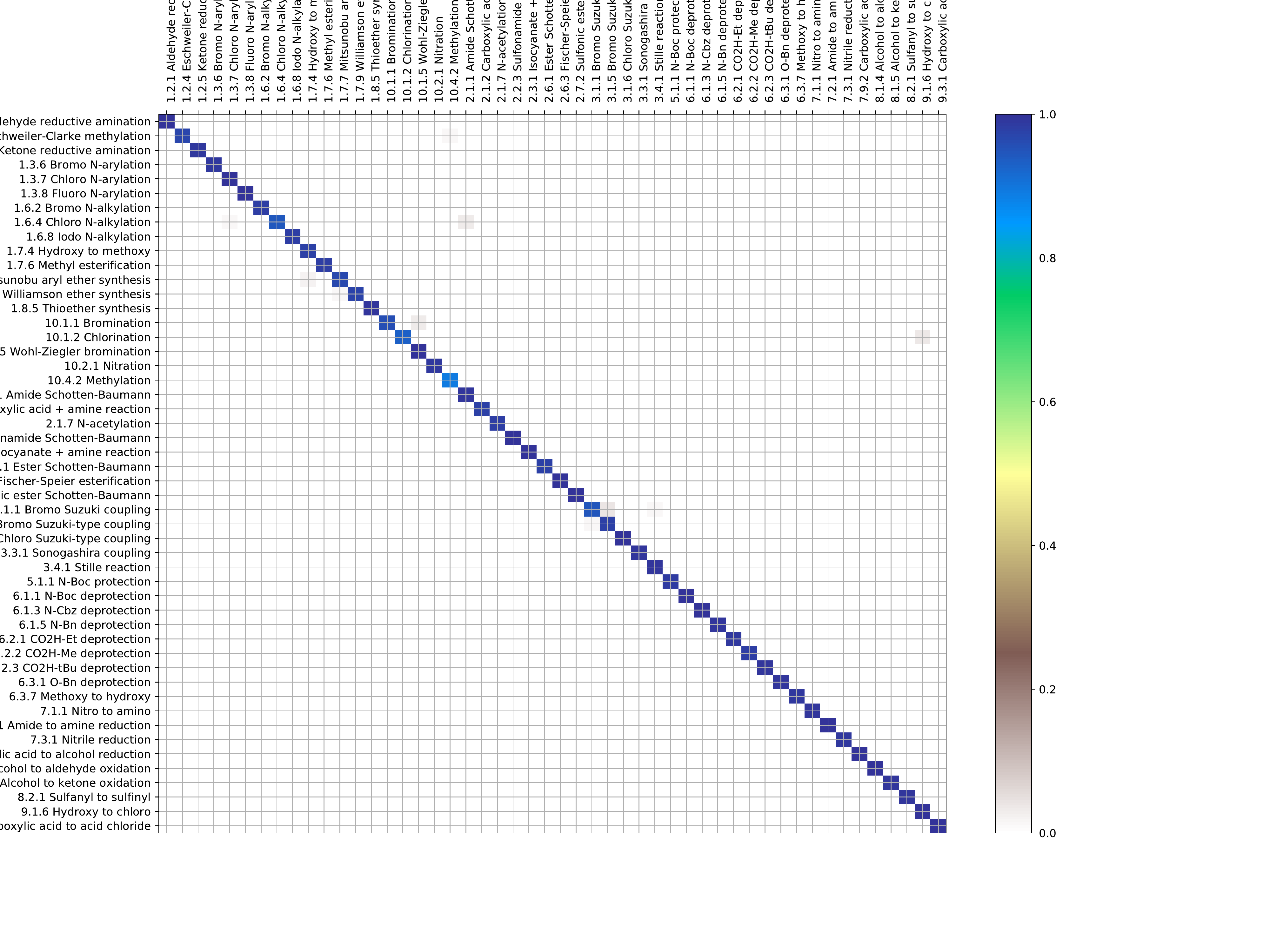}
  \caption{Confusion matrix for \emph{rxnfp (10k)} train}
  \label{fig:cm_10k}
\end{figure}

\begin{figure}[ht!]
  \centering
   \includegraphics[width=\linewidth]{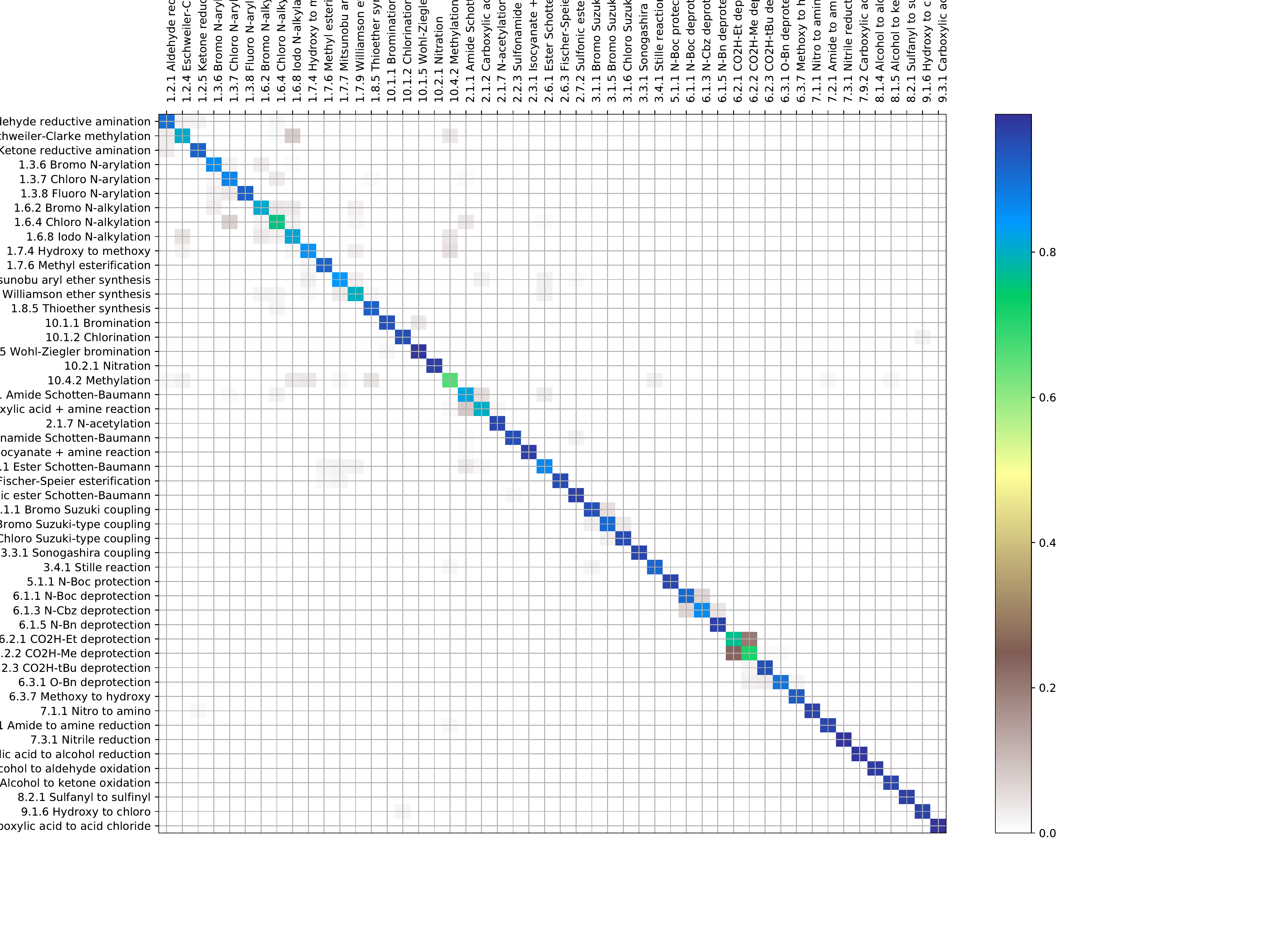}
  \caption{Confusion matrix for \emph{rxnfp (pretrained)}}
  \label{fig:cm_pretrained}
\end{figure}

 Table \ref{tab:resultsfp_10k} and Figure \ref{fig:cm_10k} show the detailed results for \emph{rxnfp (10k)}. Table \ref{tab:resultsfp_pretrained} and Figure \ref{fig:cm_pretrained} show the results of for \emph{rxnfp (pretrained)} computed by the model never fine-tuned on reaction classification. 

For both data-driven fingerprints the methylation class seems to be the hardest to predict correctly. Using the pretrained fingerprint it is hard to distinguish between reaction classes that differ only by one atom, like  ``CO2H-Et deprotection'' and ``CO2H-Me deprotection''. ``Carboxylic acid + amine condensation'' are confused with ``Amide Schotten-Baumann'' reactions and ``Mitsunobu aryl ether synthesis'' with ``Williamson ether synthesis'' reactions. It is likely that in future unsupervised reaction fingerprints will be developed that capture this fine-grained information better. 

\begin{table}[!ht]
\center
\caption{\emph{rxnfp (10k)} train: 50k reactions classification benchmark by \citet{schneider2015development} }
\footnotesize
\begin{tabular}{llllll}
\toprule
    &  recall &    prec & F-score   &   reaction class &  \\ \midrule
 0 & 0.9988 & 0.9901 & 0.9944 &Aldehyde reductive amination &1.2.1 \\
 1 & 0.9712 & 0.9848 & 0.9780 &Eschweiler-Clarke methylation &1.2.4 \\
 2 & 0.9888 & 0.9950 & 0.9918 &Ketone reductive amination &1.2.5 \\
 3 & 0.9912 & 0.9863 & 0.9888 &Bromo N-arylation &1.3.6 \\
 4 & 0.9962 & 0.9827 & 0.9894 &Chloro N-arylation &1.3.7 \\
 5 & 0.9975 & 0.9876 & 0.9925 &Fluoro N-arylation &1.3.8 \\
 6 & 0.9825 & 0.9788 & 0.9807 &Bromo N-alkylation &1.6.2 \\
 7 & 0.9437 & 0.9921 & 0.9673 &Chloro N-alkylation &1.6.4 \\
 8 & 0.9838 & 0.9825 & 0.9831 &Iodo N-alkylation &1.6.8 \\
 9 & 0.9775 & 0.9678 & 0.9726 &Hydroxy to methoxy &1.7.4 \\
10 & 0.9838 & 0.9838 & 0.9838 &Methyl esterification &1.7.6 \\
11 & 0.9675 & 0.9639 & 0.9657 &Mitsunobu aryl ether synthesis &1.7.7 \\
12 & 0.9750 & 0.9665 & 0.9708 &Williamson ether synthesis &1.7.9 \\
13 & 0.9938 & 0.9938 & 0.9938 &Thioether synthesis &1.8.5 \\
14 & 0.9575 & 0.9935 & 0.9752 &Bromination &10.1.1 \\
15 & 0.9313 & 0.9868 & 0.9582 &Chlorination &10.1.2 \\
16 & 0.9988 & 0.9685 & 0.9834 &Wohl-Ziegler bromination &10.1.5 \\
17 & 0.9888 & 0.9987 & 0.9937 &Nitration &10.2.1 \\
18 & 0.8938 & 0.9483 & 0.9202 &Methylation &10.4.2 \\
19 & 0.9950 & 0.9522 & 0.9731 &Amide Schotten-Baumann &2.1.1 \\
20 & 0.9788 & 0.9899 & 0.9843 &Carboxylic acid + amine reaction &2.1.2 \\
21 & 0.9838 & 0.9975 & 0.9906 &N-acetylation &2.1.7 \\
22 & 0.9975 & 0.9975 & 0.9975 &Sulfonamide Schotten-Baumann &2.2.3 \\
23 & 1.0000 & 0.9950 & 0.9975 &Isocyanate + amine reaction &2.3.1 \\
24 & 0.9775 & 0.9726 & 0.9751 &Ester Schotten-Baumann &2.6.1 \\
25 & 0.9962 & 0.9815 & 0.9888 &Fischer-Speier esterification &2.6.3 \\
26 & 1.0000 & 1.0000 & 1.0000 &Sulfonic ester Schotten-Baumann &2.7.2 \\
27 & 0.9463 & 0.9818 & 0.9637 &Bromo Suzuki coupling &3.1.1 \\
28 & 0.9800 & 0.9596 & 0.9697 &Bromo Suzuki-type coupling &3.1.5 \\
29 & 1.0000 & 0.9950 & 0.9975 &Chloro Suzuki-type coupling &3.1.6 \\
30 & 0.9925 & 0.9937 & 0.9931 &Sonogashira coupling &3.3.1 \\
31 & 0.9925 & 0.9778 & 0.9851 &Stille reaction &3.4.1 \\
32 & 0.9850 & 0.9975 & 0.9912 &N-Boc protection &5.1.1 \\
33 & 1.0000 & 0.9780 & 0.9889 &N-Boc deprotection &6.1.1 \\
34 & 0.9975 & 1.0000 & 0.9987 &N-Cbz deprotection &6.1.3 \\
35 & 0.9950 & 0.9925 & 0.9938 &N-Bn deprotection &6.1.5 \\
36 & 0.9888 & 0.9875 & 0.9881 &CO2H-Et deprotection &6.2.1 \\
37 & 0.9825 & 0.9800 & 0.9813 &CO2H-Me deprotection &6.2.2 \\
38 & 0.9950 & 0.9925 & 0.9938 &CO2H-tBu deprotection &6.2.3 \\
39 & 0.9950 & 0.9925 & 0.9938 &O-Bn deprotection &6.3.1 \\
40 & 0.9888 & 0.9900 & 0.9894 &Methoxy to hydroxy &6.3.7 \\
41 & 0.9938 & 0.9925 & 0.9931 &Nitro to amino &7.1.1 \\
42 & 0.9975 & 0.9803 & 0.9888 &Amide to amine reduction &7.2.1 \\
43 & 0.9912 & 0.9925 & 0.9919 &Nitrile reduction &7.3.1 \\
44 & 0.9988 & 0.9938 & 0.9963 &Carboxylic acid to alcohol reduction &7.9.2 \\
45 & 1.0000 & 0.9963 & 0.9981 &Alcohol to aldehyde oxidation &8.1.4 \\
46 & 0.9950 & 0.9987 & 0.9969 &Alcohol to ketone oxidation &8.1.5 \\
47 & 0.9950 & 0.9962 & 0.9956 &Sulfanyl to sulfinyl &8.2.1 \\
48 & 0.9962 & 0.9614 & 0.9785 &Hydroxy to chloro &9.1.6 \\
49 & 0.9975 & 0.9888 & 0.9932 &Carboxylic acid to acid chloride &9.3.1 \\ \midrule
 &  0.99 &   0.99 &   0.99 & Average & \\ \bottomrule

\end{tabular}
    \label{tab:resultsfp_10k}
\end{table}

\begin{table}[!ht]
\center
\caption{\emph{rxnfp (pretrained)}: 50k reactions classification benchmark by \citet{schneider2015development} }
\footnotesize
\begin{tabular}{llllll}
\toprule
  &  recall &    prec & F-score   &   reaction class &  \\  \midrule
 0 & 0.9012 & 0.8990 & 0.9001 &Aldehyde reductive amination &1.2.1 \\
 1 & 0.8063 & 0.8323 & 0.8190 &Eschweiler-Clarke methylation &1.2.4 \\
 2 & 0.9213 & 0.9213 & 0.9213 &Ketone reductive amination &1.2.5 \\
 3 & 0.8600 & 0.8632 & 0.8616 &Bromo N-arylation &1.3.6 \\
 4 & 0.8712 & 0.7938 & 0.8308 &Chloro N-arylation &1.3.7 \\
 5 & 0.9225 & 0.9498 & 0.9360 &Fluoro N-arylation &1.3.8 \\
 6 & 0.8113 & 0.8353 & 0.8231 &Bromo N-alkylation &1.6.2 \\
 7 & 0.7600 & 0.7696 & 0.7648 &Chloro N-alkylation &1.6.4 \\
 8 & 0.8125 & 0.7908 & 0.8015 &Iodo N-alkylation &1.6.8 \\
 9 & 0.8500 & 0.8662 & 0.8580 &Hydroxy to methoxy &1.7.4 \\
10 & 0.9200 & 0.9258 & 0.9229 &Methyl esterification &1.7.6 \\
11 & 0.8413 & 0.8519 & 0.8465 &Mitsunobu aryl ether synthesis &1.7.7 \\
12 & 0.8000 & 0.7960 & 0.7980 &Williamson ether synthesis &1.7.9 \\
13 & 0.9225 & 0.8902 & 0.9061 &Thioether synthesis &1.8.5 \\
14 & 0.9437 & 0.9461 & 0.9449 &Bromination &10.1.1 \\
15 & 0.9463 & 0.9232 & 0.9346 &Chlorination &10.1.2 \\
16 & 0.9838 & 0.9633 & 0.9734 &Wohl-Ziegler bromination &10.1.5 \\
17 & 0.9738 & 0.9725 & 0.9731 &Nitration &10.2.1 \\
18 & 0.6625 & 0.7172 & 0.6888 &Methylation &10.4.2 \\
19 & 0.8175 & 0.7861 & 0.8015 &Amide Schotten-Baumann &2.1.1 \\
20 & 0.8013 & 0.8250 & 0.8129 &Carboxylic acid + amine reaction &2.1.2 \\
21 & 0.9600 & 0.9588 & 0.9594 &N-acetylation &2.1.7 \\
22 & 0.9450 & 0.9345 & 0.9397 &Sulfonamide Schotten-Baumann &2.2.3 \\
23 & 0.9725 & 0.9569 & 0.9647 &Isocyanate + amine reaction &2.3.1 \\
24 & 0.8625 & 0.8582 & 0.8603 &Ester Schotten-Baumann &2.6.1 \\
25 & 0.9525 & 0.9658 & 0.9591 &Fischer-Speier esterification &2.6.3 \\
26 & 0.9700 & 0.9395 & 0.9545 &Sulfonic ester Schotten-Baumann &2.7.2 \\
27 & 0.9437 & 0.9333 & 0.9385 &Bromo Suzuki coupling &3.1.1 \\
28 & 0.9113 & 0.9045 & 0.9078 &Bromo Suzuki-type coupling &3.1.5 \\
29 & 0.9550 & 0.9340 & 0.9444 &Chloro Suzuki-type coupling &3.1.6 \\
30 & 0.9625 & 0.9686 & 0.9655 &Sonogashira coupling &3.3.1 \\
31 & 0.9150 & 0.9150 & 0.9150 &Stille reaction &3.4.1 \\
32 & 0.9613 & 0.9661 & 0.9637 &N-Boc protection &5.1.1 \\
33 & 0.9100 & 0.9089 & 0.9094 &N-Boc deprotection &6.1.1 \\
34 & 0.8600 & 0.9005 & 0.8798 &N-Cbz deprotection &6.1.3 \\
35 & 0.9700 & 0.9293 & 0.9492 &N-Bn deprotection &6.1.5 \\
36 & 0.7688 & 0.7437 & 0.7560 &CO2H-Et deprotection &6.2.1 \\
37 & 0.7150 & 0.7259 & 0.7204 &CO2H-Me deprotection &6.2.2 \\
38 & 0.9450 & 0.9486 & 0.9468 &CO2H-tBu deprotection &6.2.3 \\
39 & 0.8962 & 0.9459 & 0.9204 &O-Bn deprotection &6.3.1 \\
40 & 0.9313 & 0.9418 & 0.9365 &Methoxy to hydroxy &6.3.7 \\
41 & 0.9663 & 0.9898 & 0.9779 &Nitro to amino &7.1.1 \\
42 & 0.9613 & 0.9470 & 0.9541 &Amide to amine reduction &7.2.1 \\
43 & 0.9900 & 0.9888 & 0.9894 &Nitrile reduction &7.3.1 \\
44 & 0.9838 & 0.9887 & 0.9862 &Carboxylic acid to alcohol reduction &7.9.2 \\
45 & 0.9750 & 0.9750 & 0.9750 &Alcohol to aldehyde oxidation &8.1.4 \\
46 & 0.9600 & 0.9540 & 0.9570 &Alcohol to ketone oxidation &8.1.5 \\
47 & 0.9700 & 0.9898 & 0.9798 &Sulfanyl to sulfinyl &8.2.1 \\
48 & 0.9663 & 0.9748 & 0.9705 &Hydroxy to chloro &9.1.6 \\
49 & 0.9875 & 0.9925 & 0.9900 &Carboxylic acid to acid chloride &9.3.1 \\ \midrule
 &  0.90 &   0.90 &   0.90 & Average & \\ \bottomrule
 \end{tabular}
    \label{tab:resultsfp_pretrained}
\end{table}

\bibliography{bib}